\documentstyle[12pt]{article}
\begin{document}
\newcommand{\beq}{\begin{equation}}
\newcommand{\eeq}{\end{equation}}
\newcommand{\beqn}{\begin{eqnarray}}
\newcommand{\eeqn}{\end{eqnarray}}
\newcommand{\slp}{\raise.15ex\hbox{$/$}\kern-.57em\hbox{$\partial
$}}
\newcommand{\slA}{\raise.15ex\hbox{$/$}\kern-.57em\hbox{$A$}}
\newcommand{\lnA}{\raise.15ex\hbox{$/$}\kern-.57em\hbox{$A$}}
\newcommand{\lnC}{\raise.15ex\hbox{$/$}\kern-.57em\hbox{$C$}}
\newcommand{\slB}{\raise.15ex\hbox{$/$}\kern-.57em\hbox{$B$}}
\newcommand{\bP}{\bar{\Psi}}
\newcommand{\bC}{\bar{\chi}}
\title{Field-theory computation of charge-
density oscillations in a Tomonaga-Luttinger model with
impurities}
\author{
Victoria I. Fern\'andez$^{a, b}$ and Carlos M. Na\'on$^{a, b}$}
\date{March 2000}
\maketitle

\def\thepage{\protect\raisebox{0ex}{\ } La Plata 99-16}
\thispagestyle{headings} \markright{\thepage}

\begin{abstract}
\indent We compute the charge-density oscillations in a
Tomonaga-Luttinger model with impurities for a short-range
electron-electron potential.

\end{abstract}

\vspace{3cm} Pacs: 05.30.Fk\\ \hspace*{1,7 cm}11.10.Lm\\
\hspace*{1,7 cm}71.10.Pm

\noindent --------------------------------

\noindent $^a$ {\footnotesize Depto. de F\'\i sica.  Universidad
Nacional de La Plata.  CC 67, 1900 La Plata, Argentina.}

\noindent $^b$ {\footnotesize CONICET.\\ E-Mail:
victoria@venus.fisica.unlp.edu.ar\\ naon@venus.fisica.unlp.edu.ar
}

\newpage
\pagenumbering{arabic}

In a recent paper \cite{FLN} we introduced an extended version of
the Non-local Thirring model (NLT) \cite{NRT} which includes a
local interaction between the fermion current and a classical
background field $C_{\mu}$. Thus, the NLT interaction Lagrangian
density

\beq {\cal L}{_{int}}= \frac{1}{2}g^2\int d^2 yJ_{\mu}(x)V_{(\mu
)}(x,y) J_{\mu}(y), \label{1} \eeq

\noindent is now modified by adding

\beq {\cal L}_{imp}=\bar{\Psi}\lnC\Psi -M(x)\bar{\Psi}\Psi
\label{2} \eeq

The name ${\cal L}_{imp}$ refers to the fact that this density can
be used to study the interaction between electrons and impurities
\cite{KF}. Indeed, if one uses (\ref{1}) in order to describe the
forward scattering of one-dimensional spinless electrons
\cite{NRT} \cite{Voit} , then the first (second) term in the rhs
of (\ref{2}) models the forward (backward) scattering between
electrons and impurities. In particular, if we choose $C_1(x)=0$
together with

\beq C_0(x)=V \delta (x_{0}) \delta(x_{1}-d)= M(x) \label{3} \eeq

\noindent where V is a constant, our model coincides with the one
recently considered in ref.\cite{YCZC} to study Friedel
charge-density oscillations in a 1d Tomonaga-Luttinger liquid
\cite{TL} with an impurity located at $x_{1}=d$.\\

We now start by considering the general case ($C_{\mu}(x)$ and
$M(x)$ arbitrary). Using a convenient representation of the
functional delta and introducing the vector field $A_{\mu}$
(Please see ref. \cite{LN} for details), the partition function of
the model under consideration can be written as

\beq Z =  \int DA_{\mu} det(i \slp + i \gamma_{0}p_F+ g \lnA +
\lnC -M(x)) e^{-S[A]}, \label{4} \eeq

\noindent with

\beq S(A)=\frac{1}{2}\int d^2xd^2y
V_{(\mu)}^{-1}(x,y)A_{\mu}(x)A_{\mu}(y), \label{5} \eeq

\noindent where $V_{(\mu)}^{-1}$ is such that

\beq \int d^2z~ V_{(\mu)}^{-1}(z,x) V_{(\mu)}(y,z) = \delta^2
(x-y). \eeq

As shown in \cite{FLN} the partition function (\ref{4}) is
equivalent to a purely bosonic one, corresponding to a non-local
Sine-Gordon model given by

\beqn {\cal L}_{NLSG}
=\frac{1}{2}(\partial_{\mu}\varphi(x))^2+\nonumber\\ \frac{1}{2}
\int d^2y\partial_{\mu}\varphi(x)d_{(\mu
)}(x,y)\partial_{\mu}\varphi(y) + {\cal L'}_{imp}, \label{6} \eeqn
with \beqn {\cal L'}_{imp}= F_{\mu}(x)\partial_{\mu}\varphi
(x)-\frac{\alpha_0 (x)}{\beta^2}\cos \beta\varphi (x) \label{7}
\eeqn

$F_{\mu}(x)$ represents a couple of classical functions to be
related to the $C_{\mu}$'s and $\alpha_0 (x)$ is to be related, of
course, to $M(x)$. Indeed, it is straightforward to show that the
corresponding partition functions coincide provided that the
following identities hold:

\beq \frac{1} {\frac{g^2}{\pi}(p_0^2 \hat{V}_{(1)}
+p_1^2\hat{V}_{(0)}) + p^2}= \frac{\beta^2} {4\pi(p^2+
\hat{d}_{(0)} p_0^2+ \hat{d}_{(1)} p_1^2) }, \label{8} \eeq

\beq M(x) \cos (2 p_{F} x_{1})=\frac{\alpha_0 (x)}{\beta^2}
\label{9} \eeq

\noindent and

\beqn 2i\hat{V}_{(\mu)}^{-1}(-p)\hat{C}_{\mu}(-p)\frac{p_{\mu}
C(p) -2\epsilon_{\mu\nu} p_{\nu}B(p) }{\Delta (p)}-
\nonumber\\\frac{2 g^{2} p_{\mu}p_{\mu} B(p) \hat{h} (-p)}{\Delta
(p)} = \nonumber\\
 -
\beta\frac{\hat{F_{\mu}}(-p)p_{\mu}}{p^2+ \hat{d}_{(0)} p_0^2+
\hat{d}_{(1)} p_1^2} \label{10} \eeqn

\noindent where \beq A(p) = \frac{g^2}{2\pi}~ p^2 +
     \frac{1}{2}[\hat{V}_{(0)}^{-1}(p) p_1^2 +
\hat{V}_{(1)}^{-1}(p) p_0^2], \label{11} \eeq

\beq B(p) = \frac{1}{2}[\hat{V}_{(0)}^{-1}(p) p_0^2 +
           \hat{V}_{(1)}^{-1}(p) p_1^2],
\label{12} \eeq

\beq C(p) = [\hat{V}_{(0)}^{-1}(p) - \hat{V}_{(1)}^{-1}(p)] p_0
p_1, \label{13} \eeq

\beq \Delta (p) = C^2(p)-4A(p)B(p), \label{14} \eeq

\noindent and

\beq D(p,x_i,y_i) =\sum_i (e^{ipx_i}-e^{ipy_i}). \label{15} \eeq

As stated in the introductory paragraph, the path-integral
identification depicted above allows to make contact with recent
descriptions of 1d electronic systems in the presence of fixed
(not randomly distributed) impurities \cite{YCZC}. We shall now
discuss this possibility. To be specific we shall consider the
charge-density expectation value in a Tomonaga-Luttinger
electronic liquid ($\hat{V}_{(1)}=0$), with a non-magnetic
impurity located al $x_{1}= d$,

\beq \rho (x)= \langle \frac{i}{\sqrt{\pi}} \partial_{1} \varphi -
\cos (2 p_{F} x_{1}) \cos ( 2 \sqrt{\pi} \varphi ) \rangle _{{\cal
L}_{Bos}} \label{16} \eeq

\noindent where

\beqn
\begin{array}{cl}
{\cal L}_{Bos}=&\frac{1}{2}(\partial_{\mu}\varphi(x))^2+ \int
d^2y\partial_{1}\varphi(x)\frac{V_{(0)}(x,y)}{\pi}\partial_{1}\varphi(y)\nonumber\\
~&-2 i \int d^2y  p_F  x_1 ( V_{(0)}(x,y)
\partial_{1}^y)\partial_{1} \varphi (y) \nonumber\\
~&+\frac{i}{\sqrt{\pi}} C_{0}(x) \partial_{1} \varphi \nonumber\\
~&- C_{0}(x) \cos (2 p_F x_1 ) \cos (2 \sqrt{\pi} \varphi).
\end{array}
\label{17} \eeqn

It is convenient to have an even Lagrangian density. This is
easily achieved through a translation in the field $\varphi$. The
result is

\beq
\begin{array}{cl}
{\cal L'}_{Bos}=&\frac{1}{2}(\partial_{\mu}\varphi(x))^2
\nonumber\\ ~&+\int
d^2y\partial_{1}\varphi(x)\frac{V_{(0)}(x,y)}{\pi}\partial_{1}\varphi(y)\nonumber\\
~& - C_{0}(x) \cos (2 p_F x_1 ) \cos (2 \sqrt{\pi} \varphi)
\end{array}
\label{20} \eeq

\noindent where $f$ is a classical function satisfying

\beq
\partial_{0} f = 0
\label{21} \eeq

\noindent and

\beqn &\partial_{1} f +\frac{2}{\pi}\int d^2 y V_{(0)} (x,y)
\partial^{y}_{1} f\nonumber\\ ~& + \frac{i}{\sqrt{\pi}}C_{0}
+ 2 i p_{F}\int d^2 y y_{1}
\partial^{x}_{1} V_{0} (x,y) = 0 \label{22}
\eeqn

Thus we get

\beqn \rho (x)- \frac{i}{\sqrt{\pi}} \partial_{1} f = - \cos (2
p_{F} x_{1})\cos (\sqrt{ 4 \pi} f(x_{1}))\nonumber\\ \langle \cos
(\sqrt{4 \pi}\varphi ) \rangle _ {{\cal L'}_{Bos}}, \label{23}
\eeqn

\noindent where $\cos (2 p_{F} x_{1} ) \cos (\sqrt{ 4 \pi } f
(x_{1}))$ is called the Friedel oscillation and $ A(x)= \langle
\cos (\sqrt{ 4 \pi}\varphi) \rangle_{{\cal L'}_{Bos}}$ is the
corresponding envelope.\\

\indent Since ${\cal L'}_{Bos}$ is not exactly solvable, we shall
use the well-known self consistent harmonic approximation (SCHA)
\cite{SCHA}, which consists in replacing ${\cal L'}_{Bos}$ by

\beqn {\cal
L}=\frac{1}{2}(\partial_{\mu}\varphi(x))^2+\frac{m(x)}{2}
\varphi^2+\nonumber\\\frac{1}{\pi} \int d y_{1} \partial^{x}_{1}
\varphi(x_{0},x_{1})
U(x_{1}-y_{1})\partial^{y}_{1}\varphi(x_{0},y_{1})\nonumber\\
 \label{24} \eeqn

\noindent where $m(x)$ is an arbitrary function, related to the
impurity, to be variationally determined. Also note that we have
already considered the most interesting case of instantaneous
potentials of the form $V_{(0)} (x,y) = \delta (x_{0} -y_{0} )
U(x_{1}-y_{1})$.\\

\indent Let us now compute the envelope $A(x)$ in this
approximation:

\beqn A(x)= \langle e^{-\int d^2 y J(x,y)\varphi(y)}
\rangle_{{\cal L}} \label{25} \eeqn

\noindent with

\beq J(x,y)= - 2 i \sqrt{\pi} \delta^2(x-y) \label{26} \eeq

 We are interested in the case

\beq m(x)= m \delta (x_{1} - d) .\label{27} \eeq

Going to momentum space and doing a translation in the field
$\hat{\varphi}(p) \rightarrow \hat{\phi}(p)+ a(p)$, with $a(p)$ a
classical function, we find

\beq A(x)= \exp - \frac{1}{2} \int \frac{d^2 p}{(2 \pi )^2} J(-p)
a(p) = e^{i \sqrt{\pi} a(x)} \label{28} \eeq

\noindent where $a(p)$ satisfies

\beq a(p)= \frac{- J(p)}{2 R(-p)}- \int \frac{d^2 q}{(2 \pi)^2}
m(p-q) \frac{a(q)}{2 R(-p)} \label{29} \eeq

\noindent with

\beqn R(p)&=& \frac{p^2}{2}+ p_{1}^{2}\frac{U(p_{1})}{\pi}
\nonumber\\ m(k)&=& 2 \pi m \delta(-k_{0}) e^{-i k_{1} d}
\nonumber\\ J(k)&=& - 2 i \sqrt{\pi} e^{-i k.x} \label{30} \eeqn

 As we can see, the calculation of $A(x)$ is reduced to finding the
classical function $a(p)$, that is, to solving the integral
equation (\ref{29}). This is an integral equation with degenerated
kernel whose solution is \cite{integral equation}

\beq a(p)= \frac{- J(p)}{2 R(p)}+ K(p_{0}) f (-p) \label{31} \eeq

\noindent with

\beqn f(p)&=& - \frac{m}{2} \frac{e^{i p_{1} d}}{R(p)} \nonumber\\
g(q_{1})&=& \frac{e^{i q_{1} d}}{2 \pi} \nonumber\\ K(- p_{0})&=&
\frac {- \frac{1}{2} \int ^{\infty}_{- \infty} d q_{1} g(q_{1})
\frac{J(-p_{0}, q_{1})}{R(p_{0},q_{1})}}{1 - \int ^{\infty}_{-
\infty} d q_{1} g(q_{1}) f(p_{0},-q_{1})} \label{32} \eeqn

It is straightforward to verify that $a(p)$ can be written in a
more compact way as

\beq a(p) = \frac{i \sqrt{\pi}}{R(p)} e^{-i p.x} [ 1- \frac{m e^{i
p_{1} r}I ( p_{0}, r)}{\pi + m I ( p_{0}, 0)}] \label{33} \eeq

\noindent where

\beq I ( p_{0}, r) = \int^{\infty}_{0} d q_{1} \frac{\cos( q_{1}
r)}{p_{0}^{2}+q_{1}^{2} (1 + \frac{2 U( q_{1})}{\pi})}. \label{34}
\eeq

\noindent with $r=|d-x_{1}|$.\\

So we finally obtained $A(r)$ as a functional of the
electron-electron potential:

\beq A(r)= \exp-\frac{1}{\pi} \int^{\infty}_{-\infty} d p_{0}
\left(I ( p_{0}, 0)- \frac{m I ^{2}( p_{0}, r)}{\pi + m I ( p_{0},
0)}\right). \label{35} \eeq

\noindent This is the main formal result of our work. In order to
check the validity of our formula we shall now consider a
short-range interaction $U(q_{1})= U = constant$. It is convenient
to express $A(r)$ in the form

\beqn A(r)&=& \exp\left(T + W\right)(r)\eeqn

\noindent where one has defined

\beqn  T &=&- \frac{2}{\pi} \int^{\infty}_{0} d p_{0} I( p_{0},0)
\nonumber\\ W(r)&=& \frac{m}{\pi} \int^{\infty}_{-\infty} d
p_{0}\frac{I ^{2}( p_{0}, r)}{\pi + m I ( p_{0},0)}  \eeqn

These integrals are easily performed yielding

\beqn T&=&- \frac{2}{\pi} \int^{\infty}_{0} d p_{0}
\int^{\infty}_{0} d q_{1} \frac{1}{p_{0}^{2}+q_{1}^{2} (1 +
\frac{2 U}{\pi})} \nonumber\\ &=& \ln
[\frac{\mu}{\Lambda}]^{\lambda} \eeqn

\noindent and

\beqn W&=& \frac{m}{\pi} \int^{\infty}_{-\infty} d p_{0}\frac{I
^{2}( p_{0}, r)}{\pi + m I ( p_{0}, 0)} \nonumber\\ &=& -
\frac{1}{\lambda} e^{\frac{2 \mu r}{\lambda}} E_{i} ( -\frac{2 \mu
r}{\lambda})\nonumber\\ &+& \frac{1}{\lambda} e^{\frac{m
r}{\lambda ^{2}}}E_{i} (- \frac{m r}{\lambda ^{2}}), \eeqn

\noindent where $E_{i}(x)$ is the  exponential integral function,
$\mu \rightarrow 0^{+}$ and $\Lambda \rightarrow \infty$ are
infrared and ultraviolet cutoffs respectively, and $\lambda =
\sqrt{ 1 + \frac{2 U}{\pi}}$.\\

\noindent At this point, in order to compare our results with
previous computations we consider the long distance regime defined
by $\frac{2\mu r}{\lambda } << 1 $, $\frac{m r}{\lambda ^{2}} >>
1$. We get \beq A(r) = C(\lambda,\mu,\Lambda) {\frac{\lambda}{2
r}}^{\frac{1}{\lambda}} e^{-\frac{\lambda}{m r}} \label{41} \eeq

\noindent which exactly coincides with \cite{EG} and \cite{YCZC}
under the same regime.

\vspace{1cm} {\bf Acknowledgements}

The authors are supported by Universidad Nacional de La Plata
(UNLP) and Consejo Nacional de Investigaciones Cient\'{\i}ficas y
T\'ecnicas (CONICET), Argentina. They are grateful to the
organizers of the XX Encontro Nacional de Fisica de Particulas e
Campos (Brazil) where these results were presented.

\end{document}